# WAVE-UNET: Wavelength based Image Reconstruction method using attention UNET for OCT images

Maryam Viqar*[a,b], Erdem Sahin[b], Violeta Madjarova[a], Elena Stoykova[a], Keehoon Hong[c]
[a]Institute of Optical Materials and Technologies, Bulgarian Academy of Sciences, 1113, Sofia, Bulgaria; [b]Faculty of Information Technology and Communication Sciences, Tampere University, 33720, Tampere, Finland; [c]Electronics and Telecommunications Research Institute, Daejeon 34129, Republic of Korea.

## ABSTRACT

In this work, we propose to leverage a deep-learning (DL) based reconstruction framework for high quality Swept-Source Optical Coherence Tomography (SS-OCT) images, by incorporating wavelength ($\lambda$) space interferometric fringes. Generally, the SS-OCT captured fringe is linear in wavelength space and if Inverse Discrete Fourier Transform (IDFT) is applied to extract depth-resolved spectral information, the resultant images are blurred due to the broadened Point Spread Function (PSF). Thus, the recorded wavelength space fringe is to be scaled to uniform grid in wavenumber (k) space using k-linearization and calibration involving interpolations which may result in loss of information along with increased system complexity. Another challenge in OCT is the speckle noise, inherent in the low coherence interferometry-based systems. Hence, we propose a systematic design methodology WAVE-UNET to reconstruct the high-quality OCT images directly from the $\lambda$-space to reduce the complexity. The novel design paradigm surpasses the linearization procedures and uses DL to enhance the realism and quality of raw $\lambda$-space scans. This framework uses modified UNET having attention gating and residual connections, with IDFT processed $\lambda$-space fringes as the input. The method consistently outperforms the traditional OCT system by generating good-quality B-scans with highly reduced time-complexity.

## 1. INTRODUCTION

Optical Coherence Tomography (OCT) is a micrometer-resolution imaging modality with a wide diagnostic potential in medical domains, such as ophthalmology, cardiology and dermatology owing to its non-invasive and cross-sectional imaging capabilities[1-2]. It works by measuring the interference signal between back-reflected light from the sample and from the reference arm. Amongst the Fourier-domain type of OCT systems[3], SS-OCT performs spectrally resolved acquisitions using a sweeping-laser light source. This light source sweeps over a range of wavelengths and the self-explanatory term – 'depth profile' encapsulates the depth and the corresponding reflectivity information in the Fourier space. Typically, this depth encoding spectrum is Inverse Fourier Transformed using IDFT to retrieve the A-scan. Prior to this, the fringes, acquired in the wavelength space, are mapped and transformed into the wavenumber space. This procedure is referred to as calibration or k-mapping. Most commonly used methods for the linearization of fringes in the k-space incorporate interpolations and approximations, hence, the image quality may suffer due to imperfect linearity in spatial frequency[4]. If the IDFT is directly performed on the non-linear fringes, it leads to highly blurred images. This is because the Fourier transformation directly relates the physical distance (z-depth) with the wavenumbers, not with the wavelength. The reconstruction quality deteriorates with the increasing depth[5]. This resultant depth-resolved profile depicts a broad imperfectly localized PSF. Moreover, the A-line based methods are time-consuming and may need system specific a-priori calibration information before each volume acquisition. To precisely resample in the k-space, one can implement either hardware[6-7] or software methods[8-9]. The hardware-based methods usually rely on k-clocks where the challenge is to find acquisition cards that closely align with the frequency range of the clocks. The software-based methods either perform k-resampling using interpolation methods like linear, cubic spline, etc. or they directly reconstruct by utilizing the non-uniform k-space fringe as input for non-uniform Fourier domain transformation techniques[10]. Though this eliminates the necessity of calibration, the determination of IDFT bins increases the computational complexity. Moreover, mechanical movements used for rapid tuning of the wavelength in the SS-OCT systems cause repeatability issues in wavenumber vs time curve, each time acquisition is done. Hence, another issue arises due to the synchronization of the wavelength sweep and acquisition of scans, leading to timing jitters. They are responsible for multiplicative noises, which affect the phase stability[11]. This problem was addressed with the use of calibration curves, by using reference interferometric signal recorded ahead of acquisition[12]. This was later advanced into technique where simultaneously acquired signals were incorporated for calibration purposes, where in addition to the laser they use an elastic wave excitation source[13]. But this technique imposes hefty burden on data bandwidth and demands trade-off from imaging speed and phase stability. Another issue concerned with the quality of images generated by OCT systems is speckle noise, which significantly deteriorates the image-quality. An averaging based approach can be used for suppressing the speckle noise between the adjacent B-scans, when the similarity among the scans is very high. But the time-complexity associated with a such a technique is very high and, hence, makes it unsuitable for real-time applications. Moreover, for the scenarios where the sample being observed is dynamic in nature, the movement within the sample can cause artifacts if averaging is used as a method for speckle reduction. This dynamic nature of samples makes the choice of the processing method and the associated time very crucial for OCT based applications.

*maryamviqar28@gmail.com;

The motivation of the work lies in finding the way to utilize the λ-space interferometric signal directly to generate high quality OCT scans in real-time. True to the authors knowledge, there has been no attempt in exploiting the wavelength fringes directly till date to reconstruct the OCT images in real-time using DL framework.

In this work, we investigate the image quality of the OCT B-scans, reconstructed by combining the wavelength space interference signal and DL based network. The main challenge here is to find a way to give a-priori information to the DL network, about the sweeping wavelength range and the corresponding non-linear wavenumber grid to recover the uncorrupted image. We target the neural networks for image reconstruction because of their success in several image restoration and reconstruction methods where the goal is not only to generate visually pleasant images but also to retain important details and structures[14]. The existing models mainly rely on calibration and putatively linear k-space fringes for image reconstruction, which further require noise removal methods to suppress speckle noise adding to the complexity. None of these methods exploit the feature enriched precise λ-space data directly. Specifically, we want to emphasize how information in the transformed wavelength space can be adapted by neural networks to extract features capable of generating despeckled B-scans. These techniques allow to generate B-scans directly from the wavelength space, without the need of intermediate k-mapping, calibration, or post-processing noise removal, hence, they save on the time-complexity, which is crucial in real-time imaging applications. Moreover, they help to refrain from averaging as a post-processing step, enabling the proposed method to be specifically applicable in case of dynamic samples.

## 2. METHOD

In SS-OCT, the volumetric reconstruction is performed by processing a range of A-scans. These 1-dimensional scans contain the reflectivity profile corresponding to different positions in depth (i.e. axial direction) obtained from the interferometric spectrum by scanning the laser in the wavelength space.

Taking into consideration only the cross-correlation term, let the reference arm beam reflectivity be $R$, with reference mirror placed at $d_R$, let $s$ represent the reflectivity of the sample, $d$ - the depth, $S(k)$ - the source spectrum as the function of wavenumber $k$; then the OCT interferometric signal $I(k)$ can be represented using Eq.1 as:

$$I(k) = \frac{1}{4}S(k)\{R^2 + R \times FT[s(d-d_R)] + R \times FT[s(d_R-d)] + |FT[s(d)]|^2\} \qquad (1)$$

The Inverse Fourier transform (represented as $FT^{-1}$) is performed below in Eq. 2, to extract the depth (d) dependent A-line signal $A(d)$, where $\otimes$ represents the convolution operation.

$$FT^{-1}[I(k)] = \tfrac{1}{4} FT^{-1}[S(k)] \otimes (R^2 \delta(d) + R \times s(d - d_R) + R \times s(d_R - d) + FT^{-1}\{|FT[s(d)]|^2\}) \tag{2}$$

We can rewrite Eq.2 using more compact form as:

$$FT^{-1}[I(k)] = FT^{-1}\{S(k)\} \otimes \alpha(d) = P(d) \otimes \alpha(d) = A(d) \tag{3}$$

$$\text{where, } P(d) = FT^{-1}\{S(k)\}$$

The goal is to obtain this amplitude $\alpha(d)$ in Eq. (3) of the light scattered by the sample object, that is depth-$d$ dependent. The term $FT^{-1}\{S(k)\}$ can be interpreted as the point spread function (PSF) $P(d)$ which represents the blurring extent of the target sample point in an A-line signal. $A(d)$ can be used to estimate $\alpha(d)$, assuming that the PSF is spatially localized delta function. But in practice, this is not always the case, as the PSFs can be imperfectly localized. The reason may be either non-uniformity or non-linearity in spatial frequency incurred by laser spectrum of the SS-OCT system[15]. If the non-equidistant k-space fringes are directly inverse Fourier transformed, the resultant depth-resolved profile can have substantial transform errors, constructing low quality images. These are highly blurred, resolution deteriorated images.

Practically, in systems with a sweeping light source, the sweep occurs in the time domain, which dictates the fact that the captured interferometric signal is linear in time and non-linear in k-space. So, at the output, we have $I(t)$ and need to have $I(k)$ by considering the relationship between wavenumber (k) and time (t) domains. If there exists a direct linear relationship between the two, we write as:

$$k = \alpha t + k_c, \text{ where } \alpha = \frac{\Delta k}{\Delta T} \tag{4}$$

In Eq.4, $\Delta k$ represents the wavenumber range, $\Delta T$ is the time window for the range $\Delta k$, $k_c$ is the central wavenumber, $t$ lies between $-\Delta T/2$ and $+\Delta T/2$. Using Eqs. (1) and (4), we can write:

$$I(t) = \tfrac{1}{4} S(\alpha t + k_c)\{R^2 + R \times FT[s(d - d_R)] + R \times FT[s(d_R - d)] + |FT[s(d)]|^2\} \tag{5}$$

Then we perform the IDFT on Eq (5), as follows:

$$FT^{-1}{}_t[I(t)] \propto \Gamma_0(\tfrac{1}{\alpha} f) \otimes \left(R^2 \delta\left(\tfrac{1}{\alpha} f\right) + R \times s\left(\tfrac{1}{\alpha} f - d_R\right) + R \times s\left(d_R - \tfrac{1}{\alpha} f\right) + FT^{-1}{}_t\left\{\left|FT_t\left[s\left(\tfrac{1}{\alpha} f\right)\right]\right|^2\right\}\right) \tag{6}$$

Here, subscript *t* is used to demarcate the reciprocity between time (*t*) and temporal frequency (*f*) domain while performing the transform and $\Gamma_0$ represents the IDFT of the source spectrum. Eq.6 shows that for linear relationship between time and the wavenumber, IDFT could be applied directly on the time domain signal $I(t)$. But for most swept source-based OCT systems, we have an approximately linear relation with the wavelength rather than wavenumber space as follows:

$$\lambda = \beta t + \lambda_c, \text{ where } \beta = \frac{\Delta \lambda}{\Delta T} \tag{7}$$

In Eq.7, $\lambda_c$ is the central wavelength, time *t* lies between $-\Delta T/2$ and $+\Delta T/2$. This linear relationship shown in Eq. (7), is indicative of the non-linearity between *k* and *t* as:

$$t = \frac{\Delta T}{\Delta \lambda}(\lambda - \lambda_c) = \frac{2\pi \Delta T}{\Delta \lambda}\left(\frac{1}{k} - \frac{1}{k_c}\right) \tag{8}$$

Expanding the Eq. (8) into its power series around $k_c$, we get:

$$t = \frac{2\pi}{\beta}\left[-\frac{1}{k_c}\left(\frac{k}{k_c} - 1\right) + \frac{1}{k_c}\left(\frac{k}{k_c} - 1\right)^2 + \frac{1}{k_c}\left(\frac{k}{k_c} - 1\right)^3 + \cdots\right] \tag{9}$$

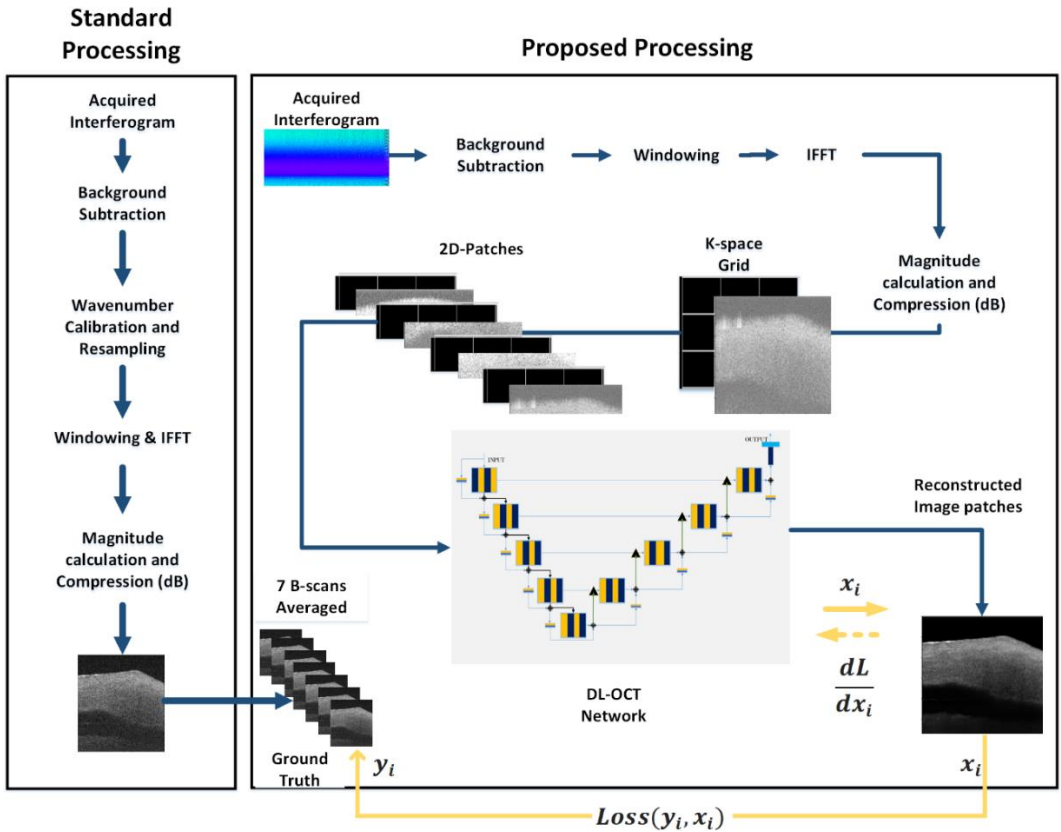

Figure 1. Standard processing in OCT, the proposed WAVE-UNET framework for OCT image reconstruction

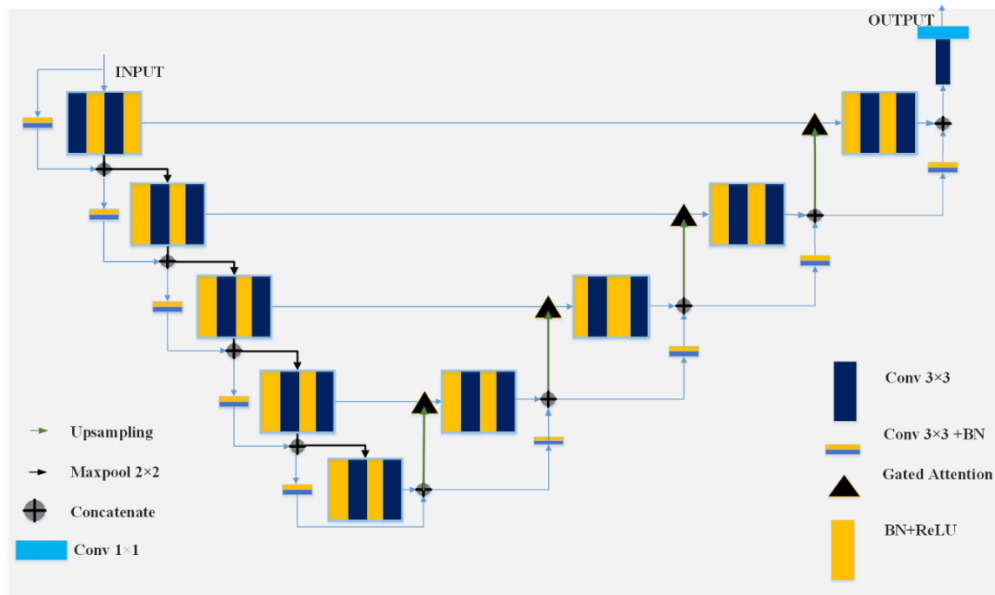

Figure 2. DL-OCT model having modified UNET with residual connections and gated attention blocks

Here, Eq. (9) represents the mathematical description of the complication in SS-OCT systems, commonly termed as non-uniform sampling in the k-space. The sequence in Eq. (9) will lead to non-linear terms in the k-space interferogram, causing loss of resolution and ranging accuracy[16]. Also, the instability and poor repeatability incurred by the laser sweeping systems create the necessity of the real-time calibration in most of the SS-OCT systems. To address this problem, we propose to use a neural network-based method to reconstruct the OCT B-scans from their spatially transformed versions using the spectral interferograms in $\lambda$-space.

We approach the high-quality image reconstruction problem through the WAVE-UNET model which utilizes the DL-OCT framework along with pre-processing steps, as shown in Fig.1. The input to this framework relies on the acquired non-uniform k-space interferometric fringes (i.e. corresponding to 1 B-scan). They are processed to get the depth resolved profile and the images which are blurred due to the broad PSF are generated. The raw $\lambda$-space fringes are pre-processed in the following manner: (i) the background subtraction of the source spectrum is performed by using the background interferometric spectrum generated in absence of the sample; (ii) the spectrum is windowed using Hann windowing; (iii) 1D IFT is applied along the vertical axis of each raw B-scan; (iv) the norm of this complex IFFT signal is calculated and converted into dB scale (complex conjugate is discarded due to conjugate symmetry). This results in an image referred to as a λ-space OCT image. This is the low-quality image owing to the fact mentioned above, that IFFT reconstruction from the non-uniform grid in wavenumber space, generates poor resolution images. The images generated as demonstrated above, are used as the input to the DL model shown in Fig.1. The DL-network incorporates the attention block with UNET[17] as the backbone architecture along with residual connections[18]. The attention blocks present at each bridge connection (between the encoder and decoder blocks) of UNET, are inspired by the work[19]. The detailed DL-OCT architecture is given in Fig. 2. In order to give a priori information about the wavenumbers, we introduce a wavenumber sharing (WS) layer. In this layer, we define a grid of wavenumbers in range: $4.62 \times 10^{6}$ rad/m - $4.99 \times 10^{6}$ rad/m associated with 1152 positions (pixels) for each A-line in a B-scan. It is interleaved as secondary layer, to the B-scan λ-space IFFT image, i.e. the intensity matrix of low-resolution input image as shown using black check design in Fig.1. This 2-dimensional input is further divided into 4 patches, where each patch is itself 2-dimensioanl, as shown in Fig.1. These patches which form the 2-layered matrix consist of layer 1 - λ-space IFFT input image and layer 2 - wavenumber layer. This 2-layered matrix is fed as input to the encoder-decoder style architecture containing 4 residual blocks on each side as the input. For each forward pass, a batch of 2-dimensioanl image-wavenumber patches is fed to the DL-network which generates the output image patch $x_i$ during the training phase. Loss is calculated between the generated output $x_i$ and the desired ground truth $y_i$. Further, using the Adam optimizer, gradients are calculated and the parameters of the model are updated accordingly. The yellow solid and dashed arrows in Fig.1 represent the forward and the backward pass of the model, respectively. Each residual block has Binary Normalization, Convolution and ReLU activation layers along with residual connections to address the gradient problem in deep architectures. The encoder down samples the input using max pooling layers while decoder upsamples the input. The skip connections between encoder and corresponding decoder blocks contain the attention gated module. The attention module[20] can improve the efficacy by getting rid of irrelevant activations, helping to focus on the details and structures.

## 3. RESULTS

The datasets used here, were acquired using the MHz Fourier domain SS-OCT (Optores GmbH, Munich) system for 5 different samples, each volume containing 600 images along with the wavelength space raw data. The OCT system uses Fourier Domain Mode Lock-FDML laser source at 1.6 MHz sweeping frequency. The central wavelength is 1309 nm with 100 nm sweeping bandwidth range. The raw data are a λ-space matrix with dimensions 2304 × 1024 points on which the image pre-processing of the framework defined in the above section is performed. We take 1152 × 1024 pixels (due to conjugate symmetry) to get the λ-space raw OCT image. The dataset overall consists of 3000 images (B-scans) obtained from 5 volumes namely varicose vein, lemon, human hand-finger, human-tooth, and cherry acquired using the SS-OCT system having resolution 1152 × 1024 (B-scan), but finally, we consider only half vertical scan 1152 × 512 due to memory constraint. Further, the patches are generated of size (288 × 512 × 2), after interleaving the wavenumber matrix to the input image and dividing this vertical image (1152 × 512 sized B-scan) into 4 patches. They were standardized by subtracting mean and dividing by standard deviation. The dataset was randomly partitioned into the percentage of 70, 20, 10 for training, validation and testing i.e. having 2100,

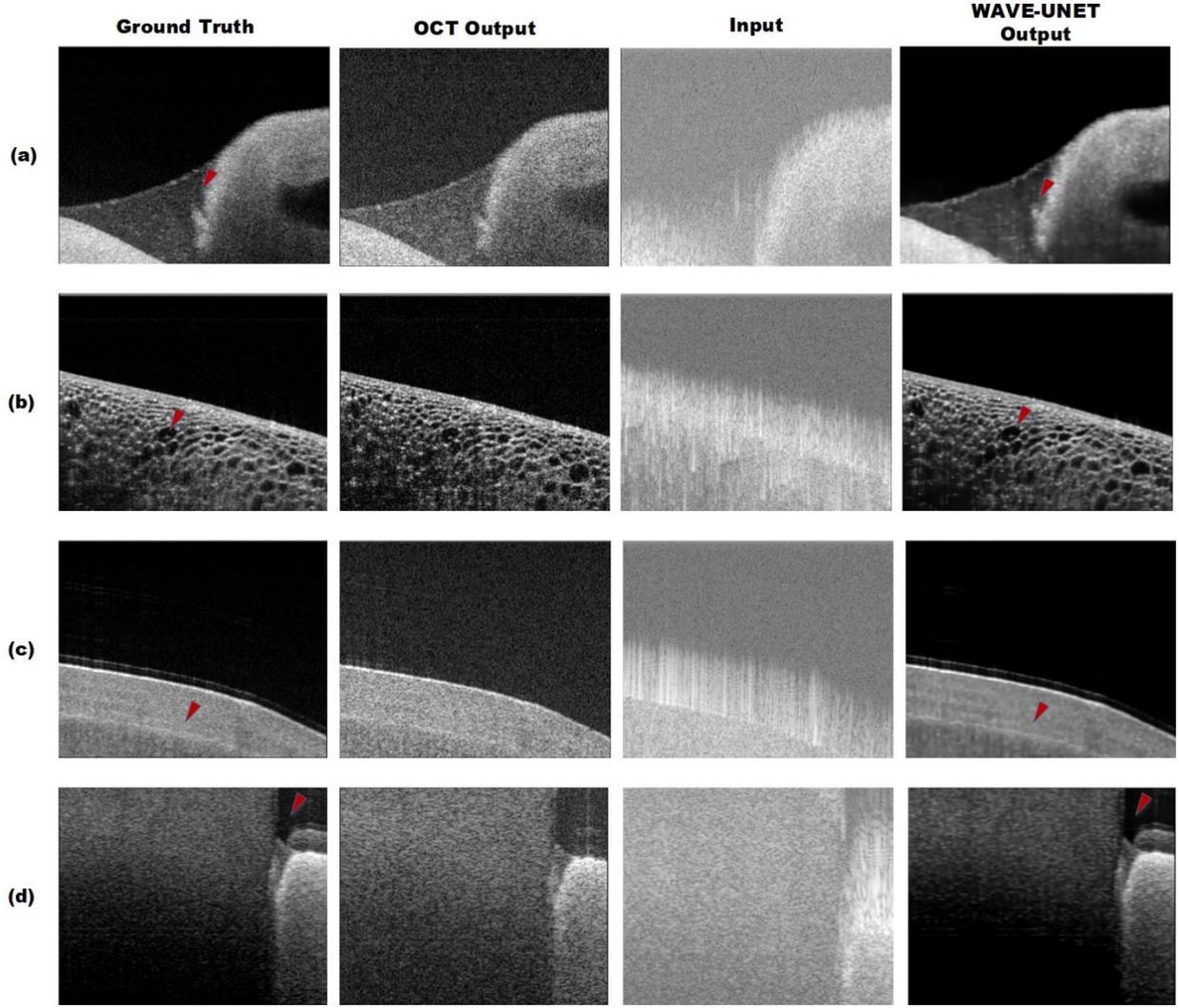

Figure 3: OCT B-scan patches from (a) vein, (b) cherry, (c) hand and (d) tooth, Columns 1 to 4 show: (1) Ground Truth (left), (2) OCT output, (3) Input Image and (4) WAVE-UNET reconstructed output image respectively

600, and 300 B-scans in each group respectively. The network was trained over 150 epochs, with a batch size of 12 patches ($288 \times 512 \times 2$). We consider the final images averaged over 7 consecutive B-scans generated by the OCT system as the ground truth for training the network. These images go through the inbuilt processing steps of k-linearization and calibration. The loss function used in the framework is mean squared error (MSE) calculated between each patch reconstructed and the corresponding ground truth.

$$MSE = \frac{1}{n}\sum_{i=1}^{n}(Y_i - X_i)^2 \qquad (10)$$

In Eq. 10, $n$ is the size of the patch (pixels), $Y_i$ represents the ground truth and $X_i$ is the predicted output at pixel $i$. The parameters were optimized using ADAM at learning rate of 0.0001. The framework was implemented on a computer with 64GB random-access memory, and an RTX 3090 graphics card. Further, we compare the ground truth and the OCT output with the results of the proposed WAVE-UNET reconstruction framework. The results obtained from the reconstruction network for an image patch of size 288 × 512 pixels are shown in Fig.3 for (a) vein, (b) cherry, (c) hand-finger and (d) tooth. It shows the ground truth (column 1), OCT output (column 2), input (column 3) and the output of the WAVE-UNET (column 4) all displayed in the intensity range 0-255. It can be clearly observed that WAVE-UNET possessed good reconstruction capability when compared with the blurred input obtained from λ-space and the OCT output which are highly corrupted with speckle noise. And the areas shown by pointing the red arrows further demarcate that it can reconstruct even fine structural details. In addition, it can be visualized that edges are also sharp, specifically in the hand and tooth outputs compared to highly blurred and noisy λ-space input scans. These features are strong enough to advocate the suitability of the proposed deep-learning based method for reconstructing high-quality images (suppressed speckle) with reduced complexity compared to images generated by the traditional pipeline used in OCT systems (without averaging).

In addition, Structural Similarity Index Metric (SSIM), Peak Signal to Noise Ratio (PSNR dB) and Mean Square Error (MSE) are used to further evaluate the network's reconstruction capability for assessing the quality of images generated. We tabulate these image quality assessment metrics in Table 1 averaged over 10 % of samples chosen randomly from each test set (i.e. vein, lemon, cherry, hand and tooth). Here I/P refers to image generated from raw λ-space (after processing), OCT refers to the image generated by OCT system using its inherent k-linearization and calibration procedures, WAVEUNET refers to the final images generated by the proposed model. In all three scenarios, the ground truth is chosen as the reference for calculating the three metrics. It can be observed, that the proposed method outperforms the original SS-OCT imaging software in terms of image quality for all the corresponding quality metrics. Significant improvement of more than 5 dB can be seen in PSNR from input towards the generated output for the lemon, vein, tooth and hand sample-set. Similarly, the SSIM increases the most for the vein sample set. Another notable fact is, for cherry, tooth, and hand, the SSIM for λ-space raw scan is higher than the OCT output image. Hence, we would also like to mention here that OCT images are highly affected by speckle and, hence, the exact SSIM values may be not directly indicative to measure the image quality independently[21]. The MSE values also follow the similar trend like in PSNR, with lemon having the lowest difference and cherry having the highest error. In all, when compared with the OCT output PSNR, SSIM and MSE, our method seems to perform substantially better as it is capable of reconstructing details and texture along with reducing speckle noise. Furthermore, it is also highly suitable for real-time processing as the inference time is 90s for a single volume containing 600 B-scans which is very low, compared to 792s for the same volume generated from the OCT system based on the standard processing, obtained on the same computer with GPU acceleration.

Table 1. Comparison of images generated by SS-OCT system and WAVE-UNET

| Metric | | VEIN | LEMON | CHERRY | TOOTH | HAND |
|---|---|---|---|---|---|---|
| **PSNR** | I/P | 17.06 | 15.41 | 15.56 | 14.57 | 14.24 |
| | OCT | 21.94 | 19.18 | 14.11 | 17.56 | 19.65 |
| | WAVEUNET | **25.50** | **27.34** | **19.21** | **19.75** | **22.81** |
| **SSIM** | I/P | 0.05 | 0.008 | 0.05 | 0.09 | 0.05 |
| | OCT | 0.21 | 0.04 | 0.03 | 0.08 | 0.03 |
| | WAVEUNET | **0.59** | **0.51** | **0.41** | **0.29** | **0.46** |
| **MSE** | I/P | 1.31 | 1.30 | 1.33 | 0.72 | 1.12 |
| | OCT | 0.42 | 0.54 | 1.86 | 0.36 | 0.32 |
| | WAVEUNET | **0.18** | **0.08** | **0.57** | **0.21** | **0.15** |

## 4. CONCLUSIONS

To summarize, in this work we propose WAVE-UNET which is a DL-based network capable of generating OCT B-scans with good reconstruction quality and reduced complexity. It rays towards a new dimension for solving inverse problems in OCT systems by using precisely linear λ-space fringes for B-scan reconstruction. It explores deep-learning based network, the UNET with residual connections and attention gates to reconstruct OCT images from blurred λ-space raw representations. This reconstruction can be done in real-time at a faster speed with reduced speckle noise, utilizing λ-space interferogram fringes bypassing intermediate steps of linearization and subsequent image processing. Consequently, this method can extract realistic structures, details and reconstruct scans that are comparable to real images. This is one of the most desirable properties in the application domain of OCT images, as their thrust area is mainly bio-medical imaging. It offers major shift of paradigm, from retaining the precise linearity of λ-space to generating good quality scans, allowing reconstruction for real-time applications. This OCT reconstruction framework has the potential to be integrated into the SS-OCT systems, to foster the acquisition speed along with the significant enhancement in signal-to-noise of generated images. Undeniably, the method needs to be experimentally tested on datasets from other OCT systems incorporating different pipelines. However, the preliminary results obtained during analysis are potent enough to be further enhanced and employed in OCT image reconstruction.

## ACKNOWLEDGMENT

MV would like to thank European Union's Horizon 2020 research and innovation programme under the Marie Skłodowska Curie grant agreement No 956770 for the funding. V.M. and E.S. thank European Regional Development Fund within the Operational Programme "Science and Education for Smart Growth 2014–2020" under the Project CoE "National center of Mechatronics and Clean Technologies" BG05M2OP001-1.001-0008. V.M. E.S. and K.H. would like to thank Institute of Information & Communications Technology Planning & Evaluation (IITP) grant funded by the Korea government (MSIT) (No. 2019-0-00001, Development of HoloTV Core Technologies for Hologram Media Services).